# A large-scale modification of Newtonian gravity as alternative approach to the dark energy and dark matter.

Falcon. N.
Universidad de Carabobo Dpto. de Física FACYT.  Venezuela. Email: nelsonfalconv@gmail.com

**Abstract:**
The peculiarities of the inverse square law of Newtonian gravity in standard Big Bang Cosmology are discussed. It is shown that the incorporation of an additive term to Newtonian gravitation, as the inverse Yukawa-like field, allows remove the incompatibility between the flatness of the Universe and the density of matter in the Friedmann equation, provides a new approach for dark energy, and enable theoretical deduce the Hubble-Lemaïtre's law. The source of this inverse Yukawa-like field is the ordinary baryonic matter and represents the large-scale contribution of gravity in accordance with the Mach's principle. It's build beginning the Matter-radiation decoupling as an specular reflection of the Yukawa potential, in agreement with astronomical and laboratory observables, result null in the inner solar system, weakly attractive in ranges of interstellar distances, very attractive in distance ranges comparable to the clusters of galaxies and repulsive in cosmic scales. Its implications in the missing mass of Zwicky, Virial Theorem, Kepler's Third Law in Globular Clusters, rotations curves of galaxies, gravitational redshift, Gravitational Lensing, the Jean's mass and BAO are discussed. It is concluded that the large-scale modification of Newtonian gravitation through the inverse Yukawa-like field allows the explicit inclusion of Mach's principle in the cosmology of the Big Bang and could be an alternative to the paradigm of non-baryonic dark matter.

**1. Introduction**.

The uncritical acceptance in cosmic scale of the Gravitational Inverse-Square Law, or Newtonian gravitation, it entails serious difficulties by to describe the dynamics of the Universe: the mass observed in the rich galaxy clusters is significantly less than the required to keep these systems stable gravitationally (the missing mass of Zwicky) and the rotation curves of the galaxies are incompatible with their virialized masses. In addition, the inconsistency between the observed average density of baryonic matter is much smaller than required by the Friedmann-Robertson-Walker models with cosmological constant (ΛFRW model) and curvature null; problem known as the missing mass. This has motivated the paradigm of non-baryonic dark matter. This hypothetical dark matter would have unknown properties, and only interacts gravitationally with ordinary matter. However, after more than a two decade of efforts: theoretical, astronomical observations and experiments, its existence has been assumed conjecturally like the ancestral paradigms of ether, phlogiston and caloric.

While it is true that the validity of inverse square law of Newton's gravity is verified with precisions greater than $10^{-8}$ for Eötvös-like experiments (Adelberger et al. 2003), there is no empirical evidence of their validity beyond the Solar System (it is assumed true for estimating the mass of binary stars). The universal character of Newton's Gravitation law was given by Kant in 1755, considering the deductive character of the planetary motion and the Leverrier's prediction for the discovery of Neptune. Recall that galaxies acquire identity after the great debate Shapley-Curtis in 1920. Also Laplace and Seeliger theorized in the eighteenth century, modifications to the Law of gravitation (Bondi 1951; Laves 1898).

On other hand, the gravitation of Newton postulates that the force of gravity has an infinite reach; even bigger than the radius of the observable Universe. Consequently implies that the rest-mass of the graviton is null, in contradiction with the Grand Unification Theories (GTU) and the detection of gravitational waves, which predict not null rest-mass (Chugreev 2017).

Another biggest problem in the hot Big Bang cosmology, closely linked to the expansion of the Universe, is the evidence of the accelerated expansion, commonly referred as dark energy, whose understanding is still unfinished (Huterer & Shafer 2018; Debono & Smoot 2016; Genova-Santos 2020). Experiment to detect dark energy forces using atom interferometry (Sabulsky et al. 2019) shows no evidence of new forces those results had places stringent bounds on scalar field theories that modify General Relativity (GR) on large scales, as Chameleon and Symmetron theories of modified gravity.

It is valid to ask if there are theoretical alternatives to Newton law of the gravity valid to large scale, concomitant with the astronomical observations and with the ΛFRW models; in the standard cosmology of the hot Big Bang.

An alternative to solve both problems: dark energy and non-baryonic dark matter, are the theories of Modified Newtonian Dynamics (MoND) (Debono & Smoot 2016), such as non-local gravitation, establishing that the force of gravitation would be the result of two term generated by ordinary matter: a first term as Newton's law, and an additional term, representing the long-range contribution in the sense of the Mach (Falcon 2013).

Mach's principle postulates that local inertia is determined by the mass distribution of the rest of the Universe (Mach 1893). In the book "The Science of Mechanics", Ernst Mach concludes that the local inertial frame is determined in some way by the movement of distant astronomical objects. This conclusion, known as Mach's Principle, arises from the very exact coincidence of the two ways of measuring the rotational speed of the angular velocity of the Earth: dynamically (Foucault's Pendulum in North Pole) and astronomically (with respect to the "Fixed stars"). This coincidence, together with Newton's well-known "bucket experiment", according to which the curvature of a rotating bucket with water is an expression of inertia ("*vis ínsita*" in Newton's words).

At any point in space, say, in the solar neighborhood, within our Galaxy or at other point in the Universe, the effective gravitational field must be the sum of the fields produced by all the celestial bodies on that point. Thus each star must provide a small contribution of the total gravitational field. Although each term looks insignificant, its sum does not have to be null, and it will exert a total gravitational field not prescribed by Newton's gravitation Law, when considering the gravitational field between two particles. Similarly, any point of space in the adjacencies of the Local Group of galaxies, will be subjected to the gravitational interaction of the baryonic mass, corresponding to the galaxy clusters and the large-scale structures. As a result, there must be a global gravitational interaction that adds to the force that mediates between any galaxies pair.

Clearly, it is not possible to calculate explicitly that global contribution to the gravitational force between two particles. Einstein tried it, through the cosmological term $\Lambda$, but it remained pending how to model its equivalent in distance ranges stellar inside a particular galaxy, and to the interior of the galaxy clusters. But it is possible to propose alternatives for modeling the term of global inertia, heuristically constructed from the observations, to represent the gravitational contribution of the large-scale distribution of baryonic matter in the local universe. It is worth asking, then, if any modification of the Law of Gravitation is possible that on a large scale solves the dynamics observed in the universe (including cosmic acceleration: Dark energy), and also, agrees with the certainties of Newtonian gravitation. Such a modification of gravitation should agree, at the scale of the solar system, with the inverse-square law and be compatible with terrestrial experiments (Eövos-like experiment).

The model of an Inverse Yukawa-like field to explicitly incorporate Mach's principle is presented in section 2, showing that it includes as a particular case the local Newtonian gravity, the theories MoND type Milgrom-like and the massive graviton. Then in section 3 the results of its inclusion in the $\Lambda$FRW-cosmology are analyzed, highlighting the solution to the problem of the Dark matter and the theoretical deduction of the Hubble-Lemaître's Law and comprehensive interpretation of dark energy. The age of the universe is also presented. Later in section 4, we discuss the Virial Theorem and the solve the Zwicky's paradox, the large-scale variation of Kepler's third Law in globular clusters, the gravitational redshift, Gravitational Lensing, the Jean's mass and BAO are discussed. Finally the conclusions are shown in the last section.

**2. The Model and Overview.**

**2.1 Phenomenology and physical argument.**

We assume that any particle with nonzero rest mass is subject gravitational inverse-square law, plus an additional force that varies with distance, caused by Inverse Yukawa-like field ($U_{YF}$). Thus the net force of gravitation varies as the law of inverse square in negligibly small distance scales in comparison to the interstellar scale, but it varies in a very different way when the comoving distance is about of the order of kiloparsec or more. In this sense, our argument is a MoND theory or large-scale modification of the Newtonian gravity.

The origin of this field ($U_{YF}$) is the baryonic mass, like in the Newtonian gravity, and represents the contribution of the gravitational distribution at great scale of the baryonic mass.

This potential per unit mass (in units of J/kg) as function of the comoving distance, is heuristically build from a specular reflection of the Yukawa potential (Falcon 2021, 2013; Falcon 2011): null in the inner solar system, weakly attractive in ranges of interstellar distances, very attractive in distance ranges comparable to the clusters of galaxies and repulsive to cosmic scales

$$U_{YF}(r) \equiv U_0(M)\,(r-r_0)\,e^{-\alpha/r} \qquad (1)$$

Where $r$ is the comoving distance in Mpc, $U_0(M) \equiv 4\pi l\, GM\, r_0^{-1}$ is an the constant, $l \equiv 1\ m^{-1}$ is a dimensional parameter.

The coupling constants; selected from the astronomical data; are: $r_0$~50 Mpc (the average distance between clusters of galaxies) and α~2.5 Mpc. We understand that an exact model could will fit the precise values of the coupling constants without modifying the phenomenology. The fig. 1 shows the variation of the $U_{YF}$ for different ranges of the comoving distance $r$ (in Mpc).

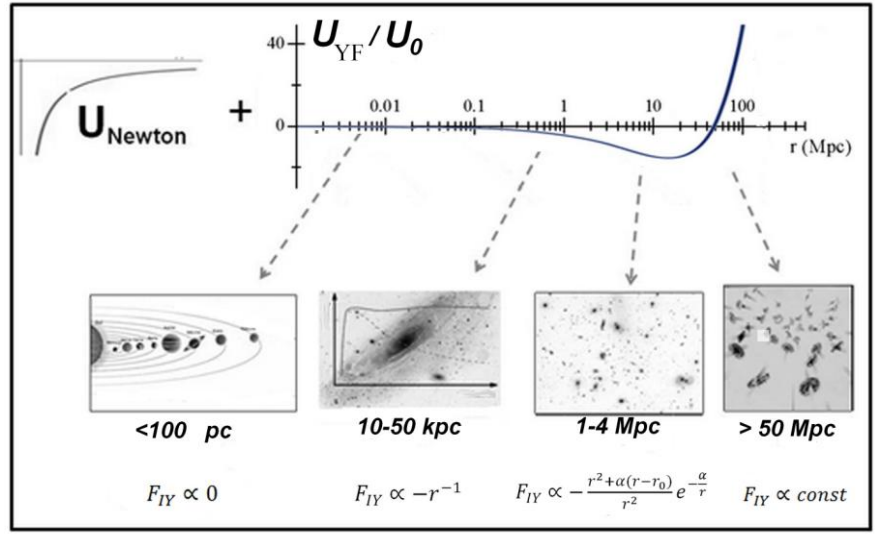

**Figure 1**: Modification of the Newtonian Dynamics with $U_{YF}$ in astronomical scale for different ranges of the comoving distance.

Note that the potential per unit mass ($U_{YF}$) results in a smooth function: continuous, differentiable, approaching zero for small range of comoving distance r. The fig. 2 show that $U_{YF}$ has a unique minimum, for the comoving distance in the order of *10 Mpc*,. In the next section, this potential is deduced from the decoupling between matter and radiation.

**2.2 Matter-radiation decoupling**

During recombination, in the first moments of the formation of the Universe, when the material is separated from radiation and hydrogen is synthesized (surface of last scattering), the average energy per unit mass (*U*) can be expressed for each nucleon a the temperature T of the plasma, using the Boltzmann distribution, as the work required to move the proton from the initial position to the comoving distance r relative to the center of the protogalactic cloud of mass *M* , as :

$$\langle U \rangle \equiv \langle \tfrac{u}{N} \rangle = -U_o(r_o - r) exp\left(-\frac{\varepsilon}{k_B T}\right) \quad (2)$$

Where $k_B$ denotes the Boltzmann constant, $U_o$ is an constant and $\varepsilon$ is the proton energy. This energy is only kinetic energy, which could be expressed in terms of the gravitational energy of the protogalactic clouds that forms later, thus

$$\varepsilon = \frac{m_p v^2}{2} \cong \frac{1}{2}\frac{G m_p M}{r} \quad (3)$$

Where, $m_p$ is the rest-mass of proton. Notice that we can use Virial theorem in its usual form, because we are at *z = 1100* and the first protostars and protogalaxies that appear much later, around *z = 6*, have not yet formed.

Using (2) and (3), then

$$U_{YF}(r) \equiv \langle U \rangle = U_0\,(r - r_0)\,e^{-\alpha/r} \quad (4)$$

With

$$\alpha = \frac{G m_p}{k_B T_d} M = \frac{4\pi G m_p}{3 k_B T_d} r_0 \rho_c \quad (5)$$

Where $T_d$ is the average plasma temperature at the moment of the decoupling ($k_B T_d \approx 13{,}6\ eV$) . M and $r_0$ is the mass and average radius of protogalaxy respectively, and $\rho_c$ is the critical density. $U_o \equiv U_o(M) = 4\pi l\ GM\ r_0^{-1}$ is a coupling constant in units of *J/kg* (l≡ 1 m$^{-1}$ is a dimensional parameter) (Falcon 2021, Falcon and Aguirre, 2014) because it is the gravitational potential in upper limit when $\varepsilon/T \to 0$ in the protogalaxy of mass $M$.

By other hand, the energy balance in the protogalaxy demands that

$$k_B T_d = \frac{G m_p}{r_0} M \tag{6}$$

Because the contribution of the large scale of gravitation ($U_{YF}$) is null in $r \approx r_0$. Remember that $U = U_N + U_{YF}$, i.e. the total gravitational energy is the Newton gravity plus the large scale contribution ($U_{YF}$-field).

Using the critical density (Falcon 2021, Falcon and Aguirre, 2014, Falcon 2013) $\rho_c \cong 9.5(8\pi)^{-1} 3 H_0^2 G^{-1} \approx 1.86 \, 10^{12} \, M_\odot / Mpc^3$ in (5), through the model of Modification of the Newtonian Dynamics, we obtain $r_0 = 47.12 \, Mpc$; and replacing this value in (4) we obtain $\alpha \cong 2.47 Mpc$. Notice that these values are closely close to the values that adjust the phenomenology of the from the astronomical data $r_0 \sim 50 \, Mpc$ and $\alpha \sim 2.5 \, Mpc$ (Falcon 2021). Thus we can write (3) with $r$ in $Mpc$, as:

$$U_{YF}(r \, in \, Mpc) \cong \mathbf{U_0} \, (r - 50) \, e^{-2.5/r} \tag{7}$$

Also, the force per unit mass (acceleration), complement to large-scale of the Newtonian gravitation is:

$$F_{IY}(r) \equiv -\vec{\nabla} U_{YF} = -\frac{U_0(M)}{r^2} e^{-\alpha/r} \left( r^2 + \alpha(r - r_0) \right) \tag{8}$$

Note that $F_{IY}$ is null by $r_c$

$$r_c = \frac{\alpha}{2} \left( \sqrt{1 + \left(\frac{4 r_0}{\alpha}\right)} - 1 \right) \cong 9.62 Mpc \approx 10 Mpc \tag{9}$$

as the previous report ) (Falcon 2021, Falcon and Aguirre, 2014), according to the assumption to the finite range of the gravitation; i.e. with massive graviton:

$$m_g^0 \cong \frac{\hbar}{r_c c} \approx 10^{-29} \frac{eV}{c^2} \tag{10}$$

And the maximum $r_m$ occurs in

$$r_m = \frac{\alpha \, r_0}{2 r_0 + \alpha} \cong 1.203 Mpc \approx 1.2 Mpc \tag{11}$$

This value is in order of the Abell radius, for the typical clusters of galaxies; as the previous assumption. We can see in (8), that if r is negligibly small compared to $r=2.5 \, Mpc$, then $F_{IY} \approx 0$, i.e. the gravity interaction is only prescribed by the inverse square law of gravitation, in accordance with Eövos-like experiments.

For ranges the comoving distances, between objects gravitationally bound, with r smaller we obtain $F_{IY} \sim r^{-1}$, see for details [5], therefore, the MoND-Milgrom results are recovered as a particular case, and could be applied to solve the problem of the rotation curves of galaxies.

## 2.3 The Eövos-like experiments and MoND -Milgrom .

Thus, using (1), the force per unit mass (acceleration), complement to large-scale of the Newtonian gravitation is:

$$F_{IY}(r) \equiv -\frac{U_0(M)}{r^2} e^{-\alpha/r} \left( r^2 + \alpha(r - r_0) \right) \tag{12}$$

That can be written in dimensionless variable as:

$$F_{IY}(x) \equiv -U_0(M)\, e^{-\alpha_0/x}\left[\frac{x^2 + \alpha_0(x-1)}{x^2}\right] \tag{13}$$

Where $x \equiv r/r_0$, $\alpha_0 \equiv \alpha/r_0 = 0.05$

Note that the maximum occurs by $x_m=0.024$ as in the core of the Abell radius, for the typical clusters of galaxies, i.e. $r_m \approx 1.2$ Mpc. (figure 2). Also note that the force is zero at $x=0.2$; that is:

$$r_c = \frac{\alpha}{2}\left(\sqrt{1+\left(\frac{4r_0}{\alpha}\right)} - 1\right) = 10\, Mpc \tag{14}$$

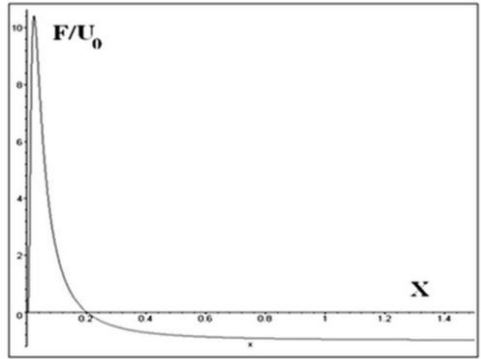

**Figure 2**: Force $F_{IY}$ in dimensionless scale $x$

. Notice that if $r$ is negligibly small compared to $\alpha=2.5$ Mpc then $F_{IY}$ is null, and the gravity is only prescribed by the inverse square law of Newtonian gravitation, in accordance with Eövos-like experiments.

For ranges the comoving distances, between objects gravitationally bound, with $r$ smaller we obtain

$$|F_{IY}(r \ll r_0)| \approx \frac{U_0(M)r_0}{2r+\alpha} \approx \left(\frac{U_0(M)r_0}{2}\right) r^{-1} \tag{15}$$

Thus $F_{IY} \sim r^{-1}$ recovers the MoND-Milgrom results as particular case. Milgrom (1983, 1983b) proposed a phenomenological modification of Newton's law which fits galaxy rotation curves solving the galaxy rotation curves problem and recovers the Tully-Fischer law. We get the result the Milgrom according to which the gravitational force depending as $r^{-1}$ (Debono & Smoot 2016, Scarpa 2006). Details of this discussion will be presented in section 4.4.

.The maximum value of the $F_{IY}$ is in $r_m \approx 1.2$ Mpc as in core of the Abell radius for the clusters of galaxies. For the average value of smooth transition to strong agglutination in galaxy's distribution ($r_c \sim 10$ Mpc) (Peebles & Rastra 2003) the $F_{IY}$ is null. From Fig. 1 it is clear that U(r) gives a constant repulsive force per unit mass, at cosmological scales providing an asymptotic cosmic acceleration. This cosmic acceleration, on a large scale, remains constant as it is observed when taking the limit of x very large, for ranges of distance like vile $x=r/r_0$ much greater than *50 Mpc*, in (3); as shown in the figure 2.

As example, in figure 3 is plotted the effective gravitational energy, per unit mass, for various members of the local group of Galaxies. In the left panel, for very close satellite galaxies (in logarithmic scale). In the right panel, to show other notable members, of the Local Group, on a linear scale. The galaxy VV124 (UGC4879) maybe the most isolated dwarf galaxy in the periphery of the Local Group, near of the minimum of gravitational energy in accord with present description. Also M3, M33, NGC 300 and NGC55 they have a gravitational potential energy per unit mass, 100 times greater than the spheroid satellite galaxies of the Milky Way. The consequences of this, in dynamic stability and in the description of the rotation curves, is outside the scope of this communication, and would be interesting as an additional test for the approach to the large-scale modification of inverse square law of Newtonian gravity.

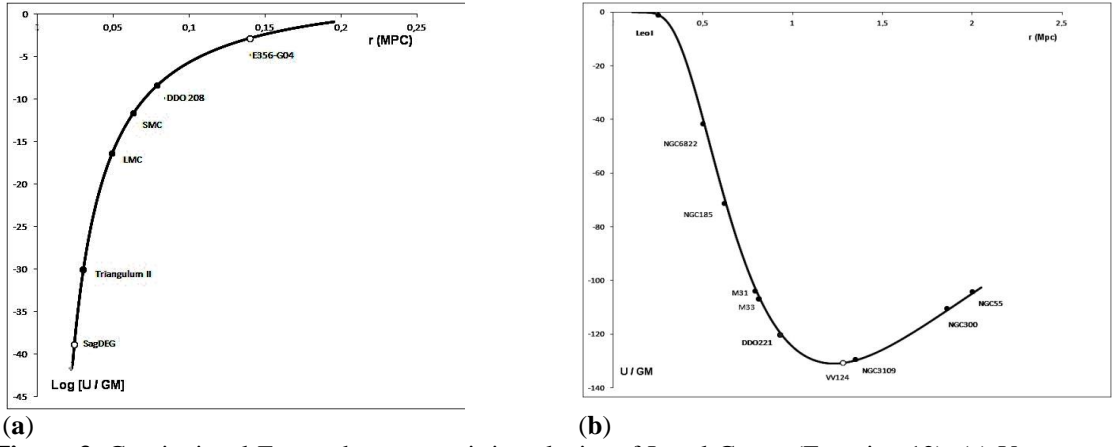

**Figure 3.** Gravitational Energy by mass unit in galaxies of Local Group (Equation 12). (a) Very near galaxies satellites of Milky Way, on logarithmic scale. (b) Galaxies of the Local group, on linear scale.

**2.4 The mass of graviton**

The value null the F$_{IY}$ in $x = 0.2$, together to the sign change in the U$_{YF}$ (fig. 1), it suggests that the range of the force of gravity is finite; i.e. in the order of 10 Mpc; and thus the graviton would have no null rest mass; then

$$m_g^0 \cong \frac{\hbar}{r_c c} \approx 10^{-64} \, kg \simeq 10^{-29} \, \frac{eV}{c^2} \tag{16}$$

Where $c$ denotes the speed of light and $\hbar$ is the Planck constant.

Massive gravity were origin in the 1930s when Wolfgang Pauli and Markus Fierz first developed a theory of a massive spin-2 field propagating on a flat space-time background (Bergshoeff et al. 2009) where the massive graviton could be decay in two photon. On 2016, the LIGO and VIRGO project reported the observational discovery of gravitational waves from a binary black hole merger event, GW150914. Their report specified a bound on the graviton mass $m_g^0 < 7.7 \cdot 10^{-23}$ eV/c² ((Bergshoeff et al. 2009 LIGO Scientific Collaboration and Virgo Collaboration 2017).
Also we can calculate the Compton wavelength associate to massive graviton as:

$$\lambda_g^0 \cong \frac{\hbar}{m_g^o c} \approx 3 \, 10^{18} \, km \tag{17}$$

In good agreement with the lower bounds inferred from the observations of binary pulsars, cluster galaxy structure and gravitational lensing (Gazeau & Novello 2011). Also the LIGO Collaboration constrains the Compton wavelength of the graviton as $\lambda_g^0 > 10^{13} \, km$ (Bergshoeff et al. 2009).

The structures found on a large scale in the distribution of galaxies and gas, with characteristic dimensions much greater than 10Mpc; e.g. Sloan Great Wall (Peebles 2020, Einasto et al. 2011) and Voids (Kovács et al. 2016); do not show symmetric axial distribution that would be expected if gravitation had infinite range. The hot gas in the superclusters of galaxies found by means of the Suyaev-Zel`dovich effect (Planck Collaboration 2011) also does not have a spherical distribution that would be expected if gravitation were of infinite range. In large-scale structures with dimensions greater than 10 Mpc, there is a gravitational bond between the galaxies, by a sequential chain of gravitational attractions between their neighboring components, but not by a common center. Assuming an infinite range for gravity, would imply among other things, to imagine colossal masses for the attractor center in the superclusters of galaxies, which are unobservable (Super Black Hole).

**2.5 Birkhoff's theorem**

The general solution to the gravitation Poisson equation, under spherical symmetry, depends on the mass distribution outside of r:

$$U(r) = -4\pi G \left[\frac{1}{r}\int_0^r \rho(r')r'^2\, dr' + \int_r^\infty \rho(r')r'\, dr'\right] \quad (18)$$

In the Newtonian gravity approach it is easily understood that the second term in (11) is canceled because of the fact that the solid angles extending from one point within a sphere to opposite directions have areas in the sphere that escalade as r2, while The gravitational force per unit the dough scales such as r-2, so that the gravitational forces of the two opposing areas are canceled exactly. The equivalent in general relativity is called Birkhoff's theorem. But in general approach, if the gravitation field has a large scale contribution, then:

$$U(r) = U_N + U_{YF} = -\frac{GM}{r} + U_0(M)(r - r_0)e^{-\alpha/r} \quad (19)$$

Thus, it`s no true that the second term in (11) are null. Therefore, Birkhoff's theorem could not be applied in the renormalized Newtonian theory of gravitation, such as in the UYF-field.

On the other hand, notice that Birkhoff's theorem says that the gravitational field outside a spherically symmetric body behaves as if all the mass of the body were concentrated in the center. But the theorem cannot be applied in general form, because its application begins by constructing 4-spheres around the center of the mass distribution, in which the gravitational field is evaluated, and in the local universe there is no way to define the center of the 4-spheres. Therefore, Birkhoff's theorem could not be applied in the renormalized Newtonian theory of gravitation, such as in the $U_{YF}$-field. Unless that we abandon the Copernican principle, suppose that the Earth is at the center of mass distribution of the Universe and then apply Birkhoff's theorem

### 3. Results: ΛFRW-Cosmology.

Let us now consider a usual ΛFRW model, with homogeneous and isotropic FRW-metric together energy-momentum tensor for a perfect fluid (Falcon 2013,Falcon 2021):

$$\mathfrak{R}^{\mu\nu} - \frac{g^{\mu\nu}}{2}\mathfrak{R} + \Lambda g^{\mu\nu} = \frac{-8\pi G}{c^2}T^{\mu\nu} \quad (20)$$

Where $\mathfrak{R}^{\mu\nu}$ is the Ricci tensor and $\mathfrak{R}$ the Ricci scalar. Without loss of generality we can write: $\Lambda \sim F_{IY}(r)$. Note that the covariance is guaranteed because at cosmological scales, where it makes sense to consider galaxies as particles of a perfect fluid, the $F_{IY}$ per unit mass is constant for ranges of the comoving distance r> 50 Mpc , as in fig 2. Thus the Dark Energy can be thought of as a "cosmic force" in the sense of the Mach Principle, caused by ordinary matter, through the Λ cosmological term.

The cosmological term leads to usual Friedmann equations (Falcon 2021):

$$\left(\frac{\dot{R}(t)}{R(t)}\right)^2 + \frac{kc^2}{R^2(t)} = \frac{8\pi G}{3}\rho + \frac{\Lambda c^2}{3} \quad (21)$$

$$\frac{2\ddot{R}(t)}{R(t)} + \left(\frac{\dot{R}(t)}{R(t)}\right)^2 + \frac{kc^2}{R^2(t)} = -\frac{8\pi G}{c^2}P + \Lambda c^2 \quad (22)$$

Now, we assumed that $\Lambda \equiv \Lambda(r) \propto F_{IY}(r)$ , as a cosmic parameter (associated to a cosmic acceleration), or dynamic variable respect to the comoving distance.

$$\Lambda \equiv \Lambda(r) = \Lambda_0 F_{IY}(r) = -\Lambda_0 \frac{d}{dr}U(r) \quad (23)$$

Where $\Lambda_0$ is the coupling constant

$$\Lambda_0 \equiv \frac{3H_0}{c^3} \cong 0.31\ 10^{-42}\ m^{-3}s^2 \quad (24)$$

## 3.1 New approach for cold dark matter

When $r \to r_m$, i.e. when $x \to 0.024$, the comic parameter $\Lambda(r)$ in the intergalactic scale, using (1) into (23), is:

$$\Lambda(r_m) = \Lambda_0 \left. F_{IY}(r) \right|_{r \to r_m} = -\Lambda_0 U_0 \left. e^{-\alpha_0/x} \left[ \frac{x^2 + \alpha_0(x-1)}{x^2} \right] \right|_{x \to 0.024} \quad (25)$$

Replacing the previous constants, we have

$$\Lambda(r_m) \simeq 10.55 (4\pi G \, kg \, m^{-2}) \frac{3H_0}{c^3} \quad (26)$$

Now, the Friedmann equations are:

$$\left( \frac{\dot{R}(t)}{R(t)} \right)^2 + \frac{kc^2}{R^2(t)} = \frac{8\pi G}{3} \rho + \frac{\Lambda(r_m) c^2}{3} \quad (27)$$

$$\frac{2 \ddot{R}(t)}{R(t)} + \left( \frac{\dot{R}(t)}{R(t)} \right)^2 + \frac{kc^2}{R^2(t)} = -\frac{8\pi G}{c^2} P + \Lambda(r_m) c^2 \quad (28)$$

Notice firstly that the definition of the critical density change, because the potential $U_{YF}$ in now not null when k=0. The critical density ($\rho_c$), using 27 and (26) is now:

$$H_0^2 \equiv \left( \frac{\dot{R}(t)}{R(t)} \right)^2 = \frac{8\pi G}{3} \rho_c + \frac{\Lambda(r_m) c^2}{3} \quad (29)$$

then

$$\rho_c = \frac{3H_0^2}{8\pi G} \left| 1 - \frac{\Lambda(r_m) c^2}{3H_0^2} \right| \cong 9.5 \frac{3H_0^2}{8\pi G} \approx 1.86 \, 10^{12} \, M_\odot / Mpc^3 \quad (30)$$

We can see the value of the critical density increases, because the critical mass has been underestimated in the usual definition. Thus, it has to take to into account the $U_{YF}$ is in addition to baryonic mass.

Using the standards notation: $\Omega_b \equiv \rho/\rho_c$, $\Omega_\Lambda \equiv \Lambda c^2 / 3H_0^2$, $q_0 \equiv -\frac{\ddot{R}}{R} H_0^{-2}$ for the density of matter, cosmological and deceleration parameters respectively; and the definition

$$\Omega_{IY} \equiv \frac{\Lambda(r_c) c^2}{3H_0^2} \quad (31)$$

The Friedmann equations 21) and (31) are:

$$\frac{kc^2}{R^2(t)} = H_0^2 \left[ \Omega_b (1 + \Omega_{IY}) + \Omega_\Lambda - 1 \right] \quad (32)$$

$$q_0 = \frac{\Omega_b}{2} (1 + \Omega_{IY})(1 + \frac{3P}{c^2 \rho}) - \Omega_\Lambda \quad (33)$$

Also, (32) and (33) can be writing as the standards $\Lambda$FRW cosmology:

$$\frac{kc^2}{R^2(t)} = H_0^2\left[(\Omega_b + \Omega_b\Omega_{IY}) + \Omega_\Lambda - 1\right] = H_0^2\left[\Omega_m + \Omega_\Lambda - 1\right] \tag{34}$$

$$q_0 = \frac{(\Omega_b + \Omega_b\Omega_{IY})}{2}(1 + \frac{3P}{c^2\rho}) - \Omega_\Lambda = \frac{\Omega_m}{2}(1 + \frac{3P}{c^2\rho}) - \Omega_\Lambda \tag{35}$$

Thus, in the present model, the $\Omega_c$ parameter of the cold dark matter would be the gravitational contribution caused by the large scale distribution of the ordinary baryonic matter. This additional contribution gravitational, incorporate the Mach's Principle through of the some heuristic representation, as the $U_{YF}$ proposed.

Replacing (25) and (31) in (34); we obtain in flat universe model (k=0):

$$1 = \Omega_b(1 + \Omega_{IY}) + \Omega_\Lambda \Rightarrow \Omega_b 11.42 + \Omega_\Lambda = \Omega_m + \Omega_\Lambda = 1 \tag{36}$$

Now, the remarkable result is that: if k=0 and $\Omega_{IY} \neq 0$ does not require the assumption of the non baryonic dark matter, neither requires exotic particles of cool dark matter, i.e. using $\Omega_b \approx 0.0223$ and $\Omega_\Lambda \approx 0.6911$ as in the CMB measurements of the Planck Collaborations (2016, 2019) we obtain $\Omega_m \approx 0.255$ and $\Omega_m + \Omega_\Lambda = 0.255 + 0.6911 \approx 1$.

## 3.2 Deduction of Hubble-Lemaître's Law

Consider the photons emitted from a remote galaxy with recession velocity $v$, and their observation in the reference local frame. Therefore, we should evaluate (1) at r >> 50Mpc, with initial condition $v = 0$ in t = 0. We find:

$$v = \int a\, dt = \int \left(\lim_{r \to \infty} F_{IY}(r)\right)\frac{dr}{c} \Rightarrow v = \int \left(\lim_{x \to \infty} F_{IY}(x)\right)\frac{dr}{c} \simeq \frac{U_0}{c} r \tag{37}$$

Replacing as before, $U_0 = 4\pi G\ kg\ m^{-2}$, we obtain de Hubble-Lemaître's Law (Falcon 2013; Falcon & Aguirre 2014):

$$V = \left(\frac{4\pi G\ell}{c}\right)r \equiv H_0 r \cong \left(86{,}3\ \frac{km}{sMpc}\right)r \tag{38}$$

It's:

$$H_0 = \frac{4\pi G\ell}{c} \tag{39}$$

Where $\ell \equiv 1\ kg\ m^{-2}$ is a dimensional parameter.

Notice that de value of $H_0$ is the theoretical upper limit, evaluate for most distant objects (r>>50 Mpc), also Falcon & Aguirre found $H_0 \approx 83.56$ km s$^{-1}$Mpc$^{-1}$ in a selected sample of 392 galaxies, in a range of 50 - 1400 Mpc (Falcon & Aguirre 2014). Earlier, Falcon & Genova-Santos (2008) found $H_0$ values as high as 73 km s$^{-1}$Mpc$^{-1}$ using the Sunyaev-Zeldovich effect and X-ray emission data in galaxy clusters, with the advantage that this method is independent of redshift.

The measurements of the anisotropies in de Cosmic Microwaves Background (CMB) constraints indirectly the set of cosmological parameters (k, $H_0$, $\Omega_m$, $\Omega_\Lambda$), by multiple statistics correlations over the acoustic peaks in the distribution of the radiation in the angular power spectrum. The results of the Planck Collaboration (2019) throw lower values for $H_0$. In the Planck Collaboration, the estimates of the cosmological parameters, show degeneration in the simultaneous estimates of $H_0$ and $\Omega_m$; Figure 3 in Planck Collaborations (2016), because the CMB measurements are not a direct measure of the Hubble constant.

Most recent direct measurements of the constant of Hubble which Space Telescope (HST), are $H_0 = 75.8^{+5.2}_{-4.9}$ km s$^{-1}$ Mpc$^{-1}$ and $78.5^{+6.3}_{-5.8}$ km s$^{-1}$ Mpc$^{-1}$ depending of the target calibration (de Jaege et al. 2020), furthermore Riess (2019) had found that $H_0 = 74.22 \pm 1.82$ km s$^{-1}$ Mpc$^{-1}$ in Large Magellanic Cloud (LMC).

In the section 3.4 show a broader discussion on the validity of the theoretical limit of H0 found here; valid for galaxies much more distant than 50 Mpc.

## 3.3 Comprehensive interpretation of dark energy

When $r \to r_c$, i.e when $x \to 0.2$, the comic parameter $\Lambda(r)$ in the intergalactic scale, using (8) into (25), is:

$$\Lambda(r_c) = \Lambda_0 \left. F_{IY}(r) \right|_{r \to r_c} = -\Lambda_0 U_0 r_0 \left. e^{-\alpha_0/x} \left[ \frac{x^2 + \alpha_0 (x-1)}{x^2} \right] \right|_{x \to 0.2} \quad (40)$$

Replacing the previous constants $\Lambda_0 U_0 r_0$, and (39) we have

$$\Lambda(r_c) \cong 0.623 \, (4\pi G \, kg \, m^{-2}) \frac{3 H_0}{c^3} = 0.623 \frac{3 H_0^2}{c^2} \quad (41)$$

Using (13) we obtain $0.623h \, 10^{-52} \, m^{-2}$ as lower limit because $F_{IY}$ is evaluated in 10 Mpc; and we obtain $10^{-52} \, h \, m^{-2} \approx 0.86 \, 10^{-52} \, m^{-2}$ as upper limit when $F_{IY}$ is evaluated in $r \to \infty$.

Now, the Friedmann equations are:

$$\left( \frac{\dot{R}(t)}{R(t)} \right)^2 + \frac{kc^2}{R^2(t)} = \frac{8\pi G}{3} \rho + \frac{\Lambda(r_c) c^2}{3} \quad (42)$$

$$\frac{2 \ddot{R}(t)}{R(t)} + \left( \frac{\dot{R}(t)}{R(t)} \right)^2 + \frac{kc^2}{R^2(t)} = -\frac{8\pi G}{c^2} P + \Lambda(r_c) c^2 \quad (43)$$

Thus, in the present model, the dark energy would be the cosmic acceleration in local frameworks, caused by the large scale distribution of the ordinary baryonic matter, as prescribed the Mach's Principle, through of $U_{YF}$ proposed. As before, replacing (40) in cosmological density parameter then:

$$\Omega_\Lambda \equiv \frac{\Lambda(r_c) c^2}{3 H_0^2} \cong 0.623 \frac{(4\pi G \, kg \, m^{-2})}{H_0 c} \approx 0.623 h^{-1} \quad (44)$$

Using (35) here, the upper limit for Hubble parameter $h=0.863$, we obtain $\Omega_\Lambda \approx 0.72$ in good agreement with the measurements of type Ia Supernovae (Peebles & Rastra 2003, Wondrak 2017; Riess et al 1998 and Perlmutter et al 1999)

## 3.4 The age of universe

Using (34) the Friedman equation (for flat universe) can be write in terms of the density cosmological parameters as:

$$H^2 = H_0^2 \left[ \Omega_m (1+z)^3 + \Omega_\Lambda \right] \quad (45)$$

The contribution of the radiation density has been omitted for simplicity, but it can be incorporated in the sum, multiplied by the factor $(1+z)^4$, without loss of generality in the discussion. In a flat Universe (k=0) we can write the limit the age at redshift z as:

$$\tau = H_0^{-1} \int_0^\infty \left[ (1+z)^3 \Omega_b (1+\Omega_{IY}) + \Omega_\Lambda \right]^{-\frac{1}{2}} \frac{dz}{z+1} \quad (46)$$

Notice that it is the $\Lambda$FRW conventional equation for the age of universe, because of arithmetic equality: $\Omega_m = \Omega_b (1 + \Omega_{IY}) = \Omega_b + \Omega_c$.

The numerical integration in the present model ($\Omega_b \approx 0.0223$, $\Omega_{IY} \approx 10.42$ $\Omega_\Lambda \approx 0.623$ and upper limit of $H_0 \approx 86,3$ km s$^{-1}$ Mpc$^{-1}$) gives that the age of the universe is $\tau \sim 11.42$ Gyr, in good agreement with the age of the White Dwarfs in the solar neighborhood (Tremblay et al 2014; Kilic et al 2017) and the 11.2 Gyr for the globular clusters in the Milky Way (Krauss et al. 2003). The direct Cosmochronology through White Dwarfs, offer an independent technique for the age of Milky Way (Tremblay et al 2014), in the inner halo the estimate age is 12 $^{+1.4}_{-3.4}$ Gyr (Kilic et al 2017) . The Planck Collaboration reports that τ ~ 13.8 Gyr (Planck Collaboration 2019). Using Planck collaboration data-set ($\Omega_b$ ≈ 0.0223, $\Omega_\Lambda$ ≈0.677 and $H_0$≈67,74) plus $\Omega_{IY}$ ≈ 10.42, we obtain τ ~ 14.43 Gyr.

## 4. Discussion:

The proposed $U_{YF}$ field falls within the finite range of gravitation and consequently a massive graviton. It is emphasized that theorizing a finite range of gravitation in a very short range of the order of 10 Mpc does not necessarily contradict the observations, since, as already mentioned, large-scale structures can be explained by successive attraction between galaxies and neighboring cluster; and they would not necessarily require an attractor center and forces of a longer range for their formation, balance and dynamics

On the other hand, we observe that Λ is constant in ranges of comoving distance much greater than 10Mpc. i.e. limits taken in (37) and (40); therefore the covariance prescribed by the Theory of Relativity are fulfiller. The $U_{YF}$-field is only locally covariant. Note further that no physical theory is globally covariant.

### 4.1 Angular Diameter distance

An observational test for the present cosmological assumption about de $U_{YF}$ is the source number count or redshift volume density distribution. The angular diameter distance is:

$$D_A = \frac{cH_0^{-1}}{z+1} \int_0^z \left[ (1+z)^3 \Omega_b (1+\Omega_{IY}) + \Omega_\Lambda \right]^{-\frac{1}{2}} dz \quad (47)$$

Where, we used (34) and the redshift dependence of universe-scale factor
.        In Fig. 4, we shown the dimensionless angular diameter with and without $U_{YF}$ (dashed lines), and only the parameter of the matter density    (continues line: Λ=0) . We can see that the $U_{YF}$ reduces the angular diameter distance for all z-values and their maximum in z≈ 2.

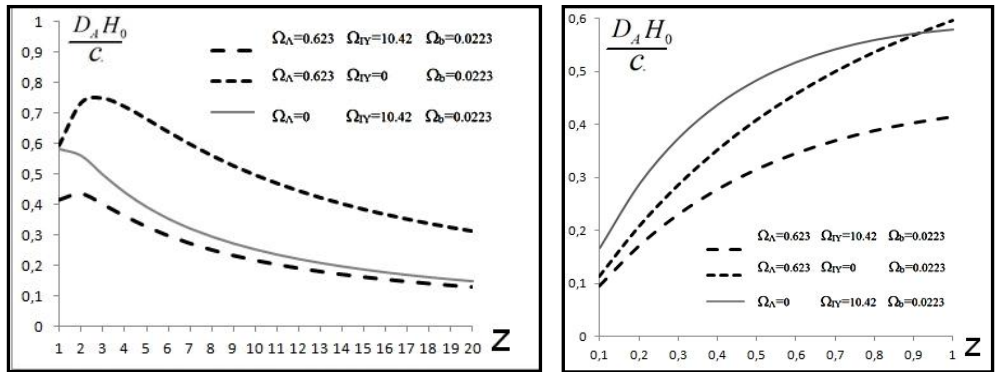

**Figure 4**: Angular diameter distance in function of the redshift for different set of cosmological parameter.

### 4.2 The Virial Theorem

In 1937 Fritz Zwicky noticed using the Virial theorem, that the Coma cluster mass are underestimated (Zwicky 1933), problem now known as the missing mass problem. Let us consider the Clausius's Virial expression $G \equiv \sum_i \vec{p}_i \cdot \vec{r}_i$ when deriving with respect to time, and then averaging with respect to a complete period (τ), we obtain the well-known virialized expression between kinetic energies and power; that in the case of a particle subjected only to the Newtonian gravitational potential is (Falcon 2012):

$$\frac{1}{\tau}\int_0^\tau \frac{dG}{dt}dt' = \frac{1}{\tau}\int_0^\tau \sum_i -\vec{\nabla} U_i . \vec{r}_i \, dt' + \frac{1}{\tau}\int_0^\tau \sum_i \frac{p_i^2}{2m_i}dt'$$

$$\frac{G(\tau)-G(0)}{\tau} = -\frac{1}{\tau}\int_0^\tau \sum_i \frac{GMm_i}{r_i}dt' + \frac{1}{\tau}\int_0^\tau \sum_i T_i dt'$$

$$0 = -\langle U \rangle + 2\langle T \rangle \qquad (48)$$

If now the particles are subjected to a gravitational potential that have an additional long-range term, we have using (1) and (8) that:

$$0 = -\langle U \rangle + 2\langle T \rangle + \frac{4\pi GM r_0^{-1} l}{\tau}\int_0^\tau \sum_i m_i e^{-\alpha/r}\left(\alpha + r_0 - \frac{\alpha r_0}{r}\right)dt' \qquad (49)$$

See the proof in Annex I. As before, $l$ is a dimensional factor with units of the kg m$^{-2}$. The integral of the last term can be evaluated using the mean value theorem, then:

$$0 = -\langle U \rangle + 2\langle T \rangle + 4\pi GM l r_0^{-1} \alpha \sum_i m_i \qquad (50)$$

Thus, the energy balance (in joules by mass unit) results:

$$\frac{v^2}{2} \equiv GM\left(-\frac{1}{r} + 4\pi l r_0^{-1} e^{-\alpha/r}(r-r_0)\right) + 4\pi \alpha l r_0^{-1} GM \qquad (51)$$

In distance ranges of galaxy clusters, the $U_{YF}$ potential dominates with respect to the Newtonian potential of the inverse square law, and then the energy balance per unit mass is:

$$\frac{v^2}{2} \cong 4\pi l r_0^{-1} GM \left[ e^{-\alpha/r}(r-r_0) + \alpha \right] \qquad (52)$$

In the particular case of the Coma cluster, with $v = 1500$ km / s and $r = 4.25$ Mpc (Shirmin 2016), thus (52) becomes:

$$GM 4\pi l r_0^{-1}\left[e^{-\alpha/r}(r-r_0)+\alpha\right]\Big|_{r=4.71} \approx 7 \cdot 10^{19} m^2 s^{-2} \gg \frac{v^2}{2} \simeq 1.12 \cdot 10^{14} m^2 s^{-2}.$$

An important result is that (50) solve the Zwicky's paradox. The "missing mass" could be interpreted as the energy associated with the $U_{YF}$ field. Annex II shows the derivation of the Tully Fischer empirical relationship using (51).

**4.3 Kepler's third law in globular clusters and rotations curves**

The deviation of the Kepler's third law in the globular clusters would be a serious test for Newtonian gravity in the outer space farther than the solar system, and would check theoretical alternatives to the non-baryonic dark matter.

The introduction of the $U_{YF}$ changes the movement equation of the astronomical bodies, and in consequence the Kepler's third law, using (8) then:

$$\frac{4\pi^2}{T^2} = \frac{GM}{r^3} - \frac{U_0(M)}{r^3}e^{-\alpha/r}\left(r^2 + \alpha(r-r_0)\right) \qquad (53)$$

Where T denote the orbital period.

The term of the Newtonian potential is greater than the term due to $U_{YF}$ for ranges less than 21.8 kpc, from which the latter becomes greater, as indicated in figure 4. In the Harrys Catalog (2010) there is at least a 10% of globular clusters with distances greater than the indicated limit value.

Note that the dynamic relationship (36) and Fig. 5 can also be used to reinterpret the rotation curves of the MoND-Milgrom theories (Falcon 2013; Milgrom 1983) where the velocity dispersion increases monotonically with radial distance instead of the Keplerian behavior predicted by the inverse law of the square of the distance.

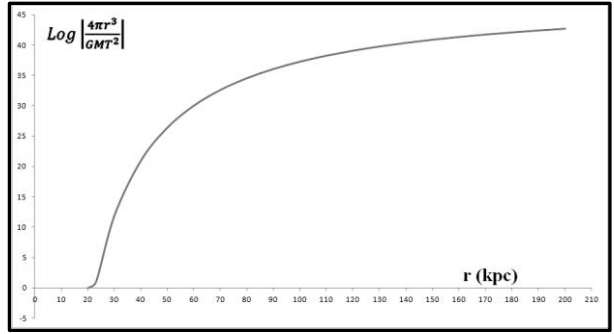

Figure 5: Expected deviation of Kepler's third law in globular Clusters as function of the comoving distance to the Milky Way center.

In Fig. 6 (left), the comparison between Newtonian forces and that associated to the $U_{YF}$ term is shown on a logarithmic scale, for interstellar distance ranges. Both forces would be comparable in the 15-40 kpc intervals, being able to explain the missing mass in the rotation curves. In the range of distances greater than 50 kpc the Newtonian force is negligible compared to the inertia caused by the large-scale distribution of matter caused by the $F_{IY}$ force.

On the other hand, in much greater distance ranges, for example for the interior of galaxy clusters, the force is proportional to the inverse of the comoving distance $F_{IY}(r) \propto (r-10)r^{-2}$, as shown in fig. 5 (right). The Newtonian force is thirty orders of magnitude smaller than $F_{IY}$ on this scale. The null value of $F_{IY}$ at $r = 10$ Mpc is in agreement with the previous discussion about the graviton's rest mass (section 2). As said before, at cosmological distance scales the $F_{IY}$ force is repulsive and manifests itself as the cosmic acceleration (Dark Energy). In ranges of comoving distances, $F_{IY}$ is negligible and the gravitational force is prescribed by the law of the inverse square of the distance (Newtonian gravitation).

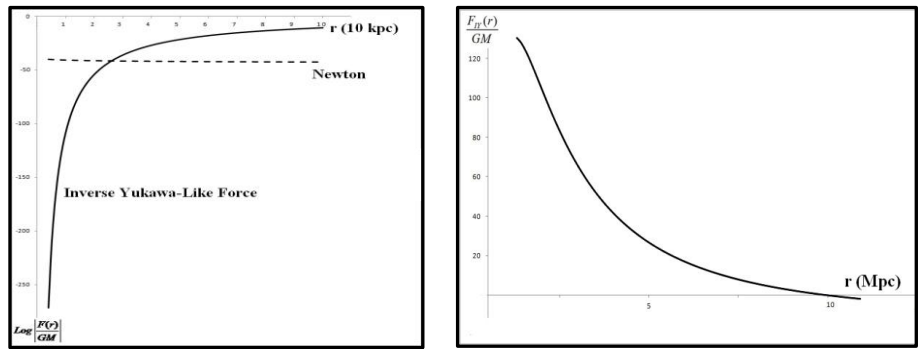

Figure 6. Relative intensity of the Force due to Yukawa's field as a function of comoving distance, in the 10-100 kpc range, for interstellar scale (left), and in the 1-10 Mpc range, for intra clusters of galaxies (right).

**4.4 Gravitational redshift.**

The total astronomical redshift is the addition of the Doppler redshift due to the emitter-receiver movement, plus the one caused by the gravitational field (gravitational redshift) and cosmological redshift due the cosmic expansion. The gravitational component of the redshift ($z_g$), in term of the gravitational potential $\phi$, (Evans & Dunning-Davies 2004) is:

$$z_g = \frac{\Delta\lambda}{\lambda_e} = \frac{\lambda_{ob} - \lambda_e}{\lambda_e} = -\frac{\phi}{c^2} \tag{54}$$

The photons emitted by the Newtonian potential source (GM / R) would also be affected by the local contribution of the gravitational field produced by the large-scale distribution of matter, then

$$z_g = -\frac{1}{c^2}\left(-\frac{GM}{R} + U_{YF}\right) \tag{55}$$

Where $R$ is the radius of the emitting source and its comoving distance is $r$. Using (1) and (13), the $U_{YF}$ can be writing as:

$$U_{YF}(x) \equiv U_0(M)\, r_0\, (x-1)\, e^{-\alpha_0/x} \tag{56}$$

Using the previous notation $x \equiv r/r_0$ and $\alpha_0 \equiv \alpha/r_0 = 0.05$; $U_0(M)\, r_0 = 4\pi GMl$, then

$$z_g = \left[\frac{R_s}{2R}\right]\left[-\frac{1}{R} + 4\pi l e^{-\alpha_0/x}(x-1)\right] \tag{57}$$

Where use the Schwarzschild's radius, as $R_s = 2GM/c^2$, and x is the comoving distance in Mpc

Now the inertial frame provides, through $U_{YF}$, a additional contribution in gravitational redshift that increase with the comoving distance ($x$), see fig.7. When the distance to the source is more less than 2.5 Mpc the additional term is null. An important question in (44) is that two galaxies an equal distance could be present redshift different in function of de $R_s$ particular of each one. These resolve the Arp's controversy (Arp 2003).

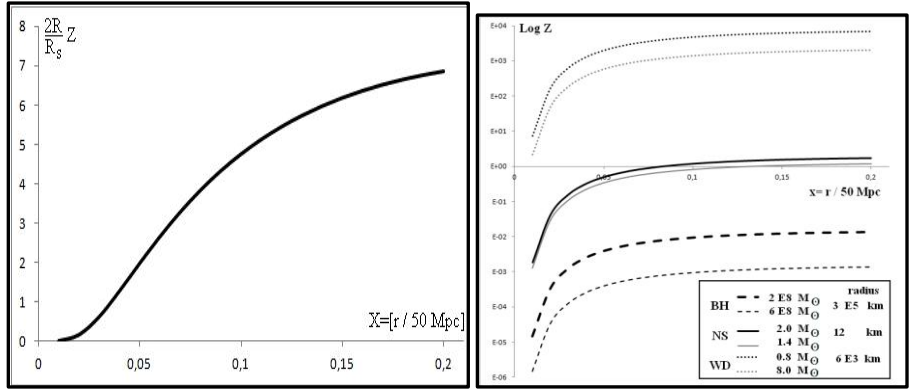

Fig.7. Gravitational redshift vs. comoving distance. For general objects in terms of the Schwarzschild radius (Left) and compact objects in terms of dimensionless $x$ (right).

Beyond 10 Mpc the gravitational redshift remains constant in accordance with the previous assumption of the massive graviton. Then the maximum additional contribution (44) of the $U_{YF}$ to the redshift gravitational is 7.83 ($x \geq 10$ Mpc) and $z_g$ increases by a factor until 88.3. This result is interesting to understand the problem of AGN at High-redshift and the nature of Quasars. Also in fig 6 (right), the variation (in logarithmic scale) of the gravitational redshift for different objects as a function of distance is shown, in particular the observation of White Dwarf's in the interior of the globular clusters could be a test for $U_{YF}$. The extreme cases of high redshift are the SBH in center of galaxies, with event horizon of the order of a light-year and masses in order of the core-mass in M87 (Event Horizon Telescope Collaboration 2019).

**4.5 Gravitational Lensing**

We have seen how the $U_{YF}$ term associated with the large-scale distribution of matter causes a change in the effective gravitational potential in the gravitational redshift; it would be expected then that it also affects the deflection of light in the formalism of gravitational lensing. Let us consider the deflection of a point-like lens of mass M, under the assumption of basic thin gravitational lent, when gravitational potential is small, the effect of the space-time curvature on light trajectory can be described as an effective refraction index η, given by:

$$\eta = 1 - \frac{2}{c^2}\Phi \cong 1 + \frac{2}{c^2}\Phi \tag{58}$$

The deflection angle (ζ) of the light rays which traveling in a gravitational field is given by the integration of the gradient component of η orthogonal to the trajectory:

$$\vec{\varsigma} = -\int \vec{\nabla}_\perp \eta \, dl = \frac{2}{c^2}\int \vec{\nabla}_\perp \Phi \, dl \tag{59}$$

But now, the gravitational field Φ is given (19), assuming that is small, i.e. α/r<<1 we obtain:

$$\Phi = U(r) = U_N + U_{YF} \cong -\frac{GM}{r} + \frac{4\pi GM\alpha_0}{r} \tag{60}$$

Where, we used $\alpha_0 \cong \alpha/r_0$, and $U_o(M) = 4\pi l \, GM \, r_0^{-1}$ as before.

If b is the impact parameter of the unperturbed light ray and $y$ denote the position along the unperturbed path as measured from the point of minimal distance from the lent: $r^2 = y^2 + b^2$,

$$\vec{\varsigma} = \frac{2}{c^2}\int \vec{\nabla}_\perp U \, dz = \frac{4GM}{c^2 b}\hat{b} - \frac{4\pi GM}{c^2 b}\alpha_0 \hat{b} \tag{61}$$

Where we using previous notation, $R_P$ denote the physical radius of stars and $R_S$ is the Schwarzschild radius, we find the deflection angle:

$$\varsigma \approx \frac{2R_S}{3c^2 R_P} \tag{62}$$

Note that the inclusion of the potential (19) leads to a reduction of one third in the calculation of the angle of deflection of the gravitational Lensing. Consequently, estimates of the deflecting mass in gravitational lenses observed at long distances (where the large-scale correction for Newtonian gravitation makes sense) would have been underestimated by a factor of three. Obviously, these do not affect the observation of the deflection of light, in the case of a total solar eclipse, because the $U_{YF}$ is null at the scale of the solar system (figure 1) because comoving range of distance is much smaller than kiloparsecs.

**4.6 Jeans mass**

In the formalism of the gravitational collapse of protogalactic clouds, the temporal evolution is considered through the gravitational amplification of density perturbation, that depend critically of the time needed for gravitational free-fall collapse ($t_g$) in compared with the travel-time of acoustic waves ($t_S$). The Jean's length ($\lambda_J^0$) is the characteristic length for which pressure balances gravity; which in term of the density (ρ), temperature ($T_e$) and hydrogen mass ($m_H$) is:

$$\lambda_J^0 \equiv \frac{c_s}{t_s} \simeq \frac{c_s}{t_g} = \sqrt{\frac{3\pi k_B T_e}{G\rho m_H}} \tag{63}$$

But the effective gravitational free-fall collapse is now due to the Newtonian potential plus the contribution of $U_{IY}$. Replacing (53) into (63), given that $tg \approx T$ then:

$$\lambda_J = \lambda_J^0 \left[1 - A\, r^{-3} e^{-\alpha/r}\left(r^2 + \alpha(r - r_0)\right)\right]^{-1/2} \tag{64}$$

Where $\lambda_J$ denotes the effective or "true" Jean's length due to the inclusion of the $U_{IY}$ field and the A parameter (Perez et al 2015), in terms of the density of the number of particles ($n_0$), it is given by:

$$A \equiv \frac{3U_o(M)}{4\pi G\rho} \simeq \frac{3Ml r_0^{-1}}{\rho} \simeq \frac{5.8 Mpc}{n_0 \, cm^3} \tag{65}$$

We observe in Fig. 8 that the modification of the Newtonian gravitation on a large scale does not affect the Jeans length in ranges of comoving distance less than 100 kpc, even for densities as low as that of the intergalactic medium. Increasing the density of the number of particles tends to smooth the difference between the Jeans lengths with and without the large-scale correction for Newtonian gravity. For low densities, such as those in the intergalactic medium, the difference between the two models for Jean's length is significant. Being of the order of 40% less than $\lambda_J^0$ in the distance ranges of (0.5 to 2) Mpc, which coincides with the agglutination of hot gas observed in X-rays in galaxy clusters.

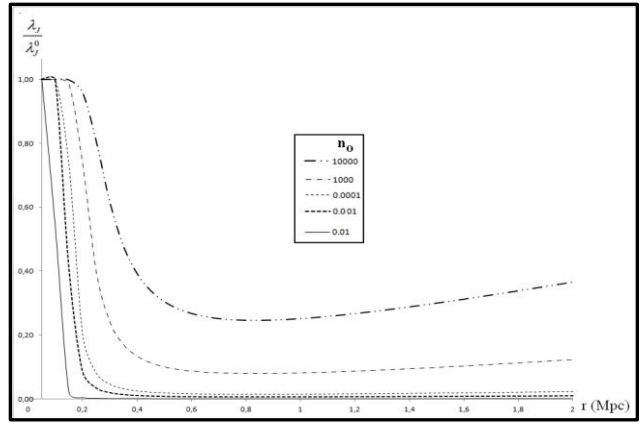

Fig.8. Variation of Jean's length at different scales.

In the previous discussion the free fall time is used, therefore it is implicitly assumed that the protostellar or protogalactic cloud has dimensions of the order of *r* in the gravitational potential (both in Newtonian and in $U_{IY}$). Both lengths coincide when x << 1 Mpc, so no changes are expected in the fragmentation of protoplanetary or protostellar molecular clouds, for which *r* << 1kpc. However, the inclusion of $U_{IY}$ does modify, as seen in figure 7, the fragmentation of protogalactic clouds; according to the initial phenomenological description (figure 1) where the dynamics is prescribed differently for the different length scales.

**4.7 BAO and CMB anisotropies**

The early universe consisted of hot plasma of photons, electrons and baryons closely coupled by Thomson scattering, with oscillations in the photon fluid, due to radiation pressure and gravity. The essential physics of these linear perturbations can be study from non-relativistic hydrodynamic approach, through equations of continuity, Euler`s and Poisson`s equations, in Lagrangian form. Now consider a small perturbation (first order) superimposed on the "background" fluid, then the fractional density perturbation (or density contrast) $\delta \cong \delta\rho/\rho_0$ obeys the relations (Peacock 1999)
:

$$\frac{d^2\delta}{dt^2} + 2H\frac{d\delta}{dt} + [c_s^2 k_s R^{-2}(t) - 4\pi G\rho_0]\delta = 0 \tag{66}$$

Where *R(t)* is the scale factor, *H* is the Hubble parameter, $c_s$ the sound speed and $k_s$ is a comoving wavenumber of a plane-wave disturbance.

Notice that the $U_{YF}$ field is not taken into account in Poisson`s equation, because the disturbances occur at the moment of decoupling long before the formation of the protostars and therefore there is not yet the distribution of masses due to the inertial frame, referred to by the Mach principle represented by the $U_{YF}$ field.

In a ΛCDM model, the growing modes of the time-dependence perturbations (Bassett and Hlozek 2009) is given by the growth function:

$$G(z) = \frac{5\Omega_m}{2} H(z) \int_z^\infty H^{-3}(z')(1+z')^3 dz' \tag{67}$$

Where the Hubble parameter is given by Friedmann equation (24), and wan be write as:

$$H(z) = \left[ (1+z)^3 \Omega_m + (1+z)^4 \Omega_{rad} + \Omega_\Lambda \right]^{1/2} \tag{68}$$

In usual formalist of growth of structure and galaxies formation, beginning by the oscillations acoustic adiabatic (BAO), it`s used $\Omega m = \Omega_b + \Omega_c$ in (37) and (38). It`s easy see that the large-scale modification of Newtonian gravitation as alternative approach to cold dark matter would given an identical result for BAO, by the arithmetic identity: $\Omega_m = \Omega_b + \Omega_c = \Omega_b(1+\Omega_{YF})$ as previous discussion in section 4.1 and 4.3, i.e. interpreting the term referring to cold dark matter without invoking the existence of non-baryonic exotic matter.

In the present $U_{YF}$-field formalist the dependence of density and pressure remain unchanged on the scale factor of expansion $R(t)$, and therefore, not affects the Early Universe. Thus, during the decoupling matter-radiation, the primordial fluctuations and the anisotropies in the CMB, the nucleosynthesis (baryogenesis), should remain unchanged by a large-scale modification of Newtonian gravity. Remark that $U_{YF}$ is forty orders of magnitude higher than the average distance per nucleon in the primordial plasma. Also the Sachs-Wolfe effect also does not change, because the size of the horizon at the time of recombination is approximately *100 kpc*, much less than the maximum range of the gravitational force with massive graviton (*~10 Mpc*) and at such ranges, the graviton would travel the entire universe inside the horizon without decay.

## 5. Conclusions

The global description of the observable Universe that we have today motivates the review of gravitation (Huterer & Shafer 2018; Debono & Smoot 2016; Genova-Santos 2020; Riess 2019, Trippe 2014) and the postulation of new theoretical alternatives. Particularly interesting could be generalizations consistent with experimental data, astronomical observations, and the well-proven formalism of existing theories. This is the case of the proposed potential $U_{YF}$, which is a heuristic construction useful to modify the Newtonian inverse-square law. The origin of this hypothetical $U_{YF}$ potential is only the ordinary baryonic matter and represents the inertia due to the distribution of matter on a large scale, incorporating the Mach principle into the formalism. The proposed potential allows a local variation of gravity depending on the range of the distance as shown in figure (1). It is consistent with the observations and leads to several important consequences:

1. All particles with no null rest mass are subject gravitational inverse-square laws, plus an additional force term that varies with distance, caused $U_{YF}$-field. At large distances from the sources, the reduction in the Newtonian field would be complemented by an additional interaction component that is grows at much greater distances.
2. This $U_{YF}$-field depend of comoving distance: null in the inner solar system, weakly attractive in ranges of interstellar distances, very attractive in distance ranges comparable to the clusters of galaxies and repulsive to cosmic scales.
3. The minimum of the potential energy of the $U_{YF}$-field, located at the order of the comoving distance of 10Mpc implies the nullity of the force of gravity, and consequently predicts a graviton mass of at least *$10^{-64}$* kg.
4. Also the Force ($F_{IY}$) supplementary to Newtonian gravity would reach a maximum for distances of the order of 1.2 Mpc, favoring the agglutination of matter in galaxy clusters, and would be evidenced as a maximum in the redshift volume density distribution around $z = 2$ (fig. 4)
5. The inclusion of the large scale term of the gravity, remove the incompatibility between the flatness of the Universe and the density of matter in the Friedmann equation, without invoking the non baryonic dark matter (34).
6. The usual cold dark matter parameter ($\Omega_c$) would be, the gravitational contribution caused by the large scale distribution of the ordinary baryonic matter thought the $U_{YF}$-Field (36).
7. The dark energy would results to be the cosmic acceleration in local frameworks, caused by the large scale distribution of the ordinary baryonic matter, as prescribed the Mach's principle, through of the $U_{YF}$ proposed (44)
8. The Hubble-Lemaître's law would be the manifestation on a cosmic scale of the $U_{IY}$ field, with theoretical expression as in (38). The theoretical upper limit, evaluated for the most distant objects ($r >> 50$ Mpc) would be 86,3 $kms^{-1}Mpc^{-1}$.
9. The age of the universe would not be much older than $\tau \sim 11.42$ Gyr, in good agreement with direct WD-Cosmochronology.
10. The Virial theorem is formally derived to show that the field $U_{YF}$ implies an additional term in the energy balance due to the distribution of matter on a large scale. Therefore the missing mass could be interpreted as the energy associated with the $U_{YF}$ field and it's solve the Zwicky's paradox.

11. The comparison between Newtonian forces and that associated to the $U_{YF}$ term results in: both forces would be comparable in the 15-40 kpc intervals, being able to explain the missing mass in the rotation curves. Inside galaxy clusters, the force is proportional to the inverse of the comoving distance (Fig. 6).
12. The $U_{YF}$ provides an additional contribution for the gravitational redshift that increases with the comoving distance (57) and $z_g$ increases by a factor until 88% ($r >> 50$ Mpc).
13. The increase in gravitational redshift due to the $U_{YF}$ field also implies that two galaxies located at the same distance could exhibit a different redshift based on their particular Schwarzschild radii. This solves the Arp's controversy. This result is interesting to understand the problem of AGN at high-redshift and the nature of quasars.
14. The inclusion of the $U_{YF}$-field leads to a reduction in the calculation of the angle of deflection of the gravitational lensing so the deflecting mass at long distances would have been underestimated by a factor of three, without affect the deflection in total eclipses of sun.
15. The modification of the Newtonian gravitation on a large scale does not affect the Jean's length in ranges of comoving distance less than 100 kpc (Fig. 8). but the Jean's length results 40% less at distance ranges of Mpc, which coincides with the agglutination of hot gas in X-rays galaxies clusters.
16. The energy density of the universe must increase due to the graviton's rest mass, which is incorporated through the $U_{YF}$ field in Friedmann's equation (30). Consequently, the critical density no longer corresponds to the Einstein-de Sitter model, which lacks a material basis.
17. The large-scale modification of Newtonian gravitation given an identical result for BAO as the paradigm of cold dark matter, by the arithmetic identity: $\Omega_m = \Omega_b + \Omega_c = \Omega_b(1+\Omega_{YF})$; but it does not require assuming the existence of non-baryonic exotic particles
18. The CMB and primordial nucleosynthesis remain unchanged by a large-scale modification of Newtonian gravity, such as the $U_{YF}$-field discussed here

Regardless of whether the expression for the so-called "Inverse Yukawa-like field" ($U_{YF}$) proposed is exactly the proposal here, we see that their inclusion could be a viable alternative to the paradigm of non-baryonic dark matter and is concomitant with Hot Big Bang model and the most recent astronomical observations; and would be thinks with the usual Physics. Perhaps, to understand the dynamics of the universe, we would have to abandon the assumption about the universal validity of the inverse square law of Newtonian gravity, assumed uncritically, instead of conjecturing unobservable exotic matter

**Annex 1 Additional note about the Virial theorem**

Beginning the Clasius's Virial Expression then:

$$\frac{1}{\tau}\int_0^\tau \frac{dG}{dt}dt' = \frac{1}{\tau}\int_0^\tau \sum_i -\vec{\nabla} U_i \cdot \vec{r}_i \, dt' + \frac{1}{\tau}\int_0^\tau \sum_i \frac{p_i^2}{2m_i}dt'$$

$$\frac{G(\tau)-G(0)}{\tau} = -\frac{1}{\tau}\int_0^\tau \sum_i \left( \frac{GMm_i}{r_i} + m_i \vec{F}_{IY}(r) \cdot \vec{r}_i \right) dt' + \frac{1}{\tau}\int_0^\tau \sum_i T_i dt$$

$$0 = -\frac{1}{\tau}\int_0^\tau \sum_i \left( \frac{GMm_i}{r_i} + \frac{m_i U_0(M)}{r} e^{-\alpha/r}\left[r^2 + \alpha(r-r_0)\right] \right) dt' + 2\langle T \rangle$$

$$0 = -\left\langle \sum_i \frac{GMm_i}{r_i} + \sum_i m_i U_0(M) e^{-\alpha/r}(r-r_0) \right\rangle + 2\langle T \rangle + \frac{1}{\tau}\int_0^\tau \sum_i m_i U_0(M)\, e^{-\alpha/r}\left( \alpha + r_0 - \frac{\alpha r_0}{r} \right)dt'$$

The term within the integral is a continuous function, whose domain are all real numbers, differentiable on the interval $(0, \tau)$ and bounded by 1 and $f(\alpha)$, then follow with (38).

---

**Annex 2 Additional note about the Tully Fischer Relation**

The comoving distance (r) into an galaxy is the order of kiloparsec, then $r-r_0 \approx -r_0$ and $\mathrm{Exp}(-\alpha/r) \cong (1-\alpha/r)$; thus using (51) we obtain:

$$\frac{v^2}{2} = GM\left|\frac{1}{r} - 4\pi l\left(1 - \frac{\alpha}{r}\right)\right| + 4\pi\alpha l r_0^{-1}GM \approx \frac{GM}{r}(1 - 4\pi l\alpha)$$

Then, the luminosity L and the absolute brightness I are proportional's to the fourth power of the orbital velocity

$$v^4 \sim \frac{M^2}{r^2} = \frac{4\pi I}{L}M^2 \propto L$$

where it has been assumed, as usual, that the Luminosity is proportional to the stellar mass